# Temporally and spatially variant-resolution illumination patterns in computational ghost imaging

Dong Zhou,[1] Jie Cao,[1,2,*] Li-xing Lin,[1] Haoyu Zhang,[1] Huan Cui,[1] YingQiang Zhang,[1] and Qun Hao[1,*]

[1] *School of Optics and Photonics, Beijing Institute of Technology, Key Laboratory of Biomimetic Robots and Systems, Ministry of Education, Beijing 100081, China*
[2] *Yangtze Delta Region Academy, Beijing Institute of Technology, Jiaxing 314003, China*
*\* ajieanyyn@163.com, qhao@bit.edu.cn*

**Abstract:** Conventional computational ghost imaging (CGI) uses light that carries a sequence of patterns with uniform resolution to illuminate an object and then performs correlation calculations based on the light intensity value reflected by the object and the preset patterns to obtain object images. CGI requires numerous measurements to obtain high-quality images, especially if high-resolution images are to be obtained. To solve this problem, we developed temporally variable-resolution illumination patterns, replacing the conventional uniform-resolution illumination patterns with a sequence of patterns of different imaging resolutions. In addition, we propose combining temporally variable-resolution illumination patterns and spatially variable-resolution structures to develop temporally and spatially variable-resolution (TSV) illumination patterns, which improve not only the imaging quality of the region of interest (ROI) but also the robustness to noise. The methods using the proposed illumination patterns are verified by simulations and experiments compared with uniform-resolution computational ghost imaging (UCGI). For the same number of measurements, the method using temporally variable-resolution illumination patterns has better imaging quality than UCGI but is less robust to noise. The method using TSV illumination patterns has better imaging quality in the ROI than the method using temporally variable-resolution illumination patterns and UCGI with the same number of measurements. We also experimentally verify that the method using TSV patterns has better imaging performance when applied to higher resolution imaging. The proposed methods are expected to solve the current CGI problem that hinders high-resolution and high-quality imaging.

## 1. Introduction

CGI, which is a unique imaging technique that uses a single-pixel detector, is achieved by illuminating light with modulated patterns onto objects and then by correlating the modulated patterns with the light intensity collected by the detector to obtain information about the object [1]. This unique imaging system enables CGI to have the advantages of scattering robustness, wide spectrum, and beyond-visual-field imaging and has been applied in many fields, including three-dimensional imaging [2-6], terahertz imaging [7-10], target tracking [11-13] and multispectral imaging [14-17].

However, CGI is inferior to conventional imaging in terms of imaging performance, mainly because performing a full measurement requires that the number of measurements is equivalent to the total number of resolutions [18-20]. Although various algorithms have been proposed to reduce the number of measurements, the time required for the reconstruction algorithm also affects the imaging efficiency. In 2018, Higham et al. realized the recovery of real-time 128×128 pixel video at 30 frames-per-second sampling at a compression ratio of 2% using deep learning with convolutional autoencoder networks [21]. In the same year, Sun et al. proposed a

CGI scheme that utilizes an LED. The proof-of-principle system achieved continuous imaging with a 1000 fps frame rate at 32×32 pixel resolution [22]. In 2021, Hahamovich et al. developed an approach for GI that relies on cyclic patterns coded onto a spinning mask and demonstrated its imaging rates of up to 72 frames per second at 101× 103 pixel resolution [23]. Increased resolution needs to be achieved at the expense of imaging efficiency. Obtaining high-resolution object images while balancing imaging quality and imaging efficiency remains an important challenge to achieving real-time and high-quality CGI.

In most CGI methods, the illumination patterns comprise a sequence of patterns with uniform resolution [24-29]. In the spatial domain, an illumination pattern with nonuniform resolution was proposed by Phillips et al. in 2017. This scheme can improve the imaging quality of high-resolution regions at the expense of information in low-resolution regions [30]. Compared with the uniform resolution scheme, the same imaging quality can be obtained with fewer measurements, which improves the imaging efficiency. In the time domain, Zhou et al. [31] proposed multiresolution progressive CGI, which uses the Hadamard-derived pattern to simply and quickly realize continuous multiresolution imaging, in 2019. This scheme can also reduce the reconstruction time and number of measurements.

In this paper, we propose a temporally variable-resolution CGI (TVCGI) method and a temporally and spatially variable-resolution CGI (TSVCGI) method. We developed and demonstrated TVCGI, which balances the imaging quality and efficiency of reconstructed images. We designed TSV illumination patterns to enhance the imaging quality of the TVCGI in the ROI and robustness to noise. We demonstrated that the proposed method is suitable for different resolutions and that better imaging performance can be achieved, especially in high-resolution applications.

## 2. Principles

The principle of conventional CGI is shown in Fig. 1. Light carrying a sequence of random illumination patterns is directed onto the surface of the object. The light reflected from the object is collected by the single-pixel detector, and the data are processed to obtain the light-intensity value that corresponds to the illumination pattern. The reconstructed image is calculated based on the random patterns and light-intensity value. There are now certain reconstruction algorithms, such as second-order correlation [32, 33], compressive sensing [18, 34, 35], deep learning [36-38], and inverse transform [39-40]. In this paper, the total variation regularization prior algorithm [28] in compressive sensing is selected for verification.

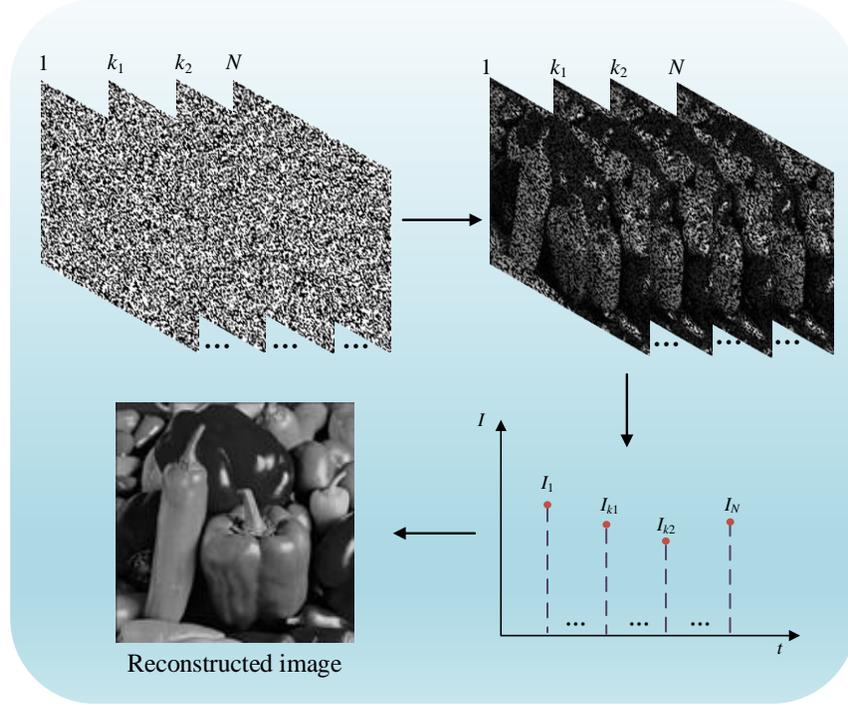

Fig. 1. Principle of conventional CGI.

The measurement principle of CGI for an object $O$ ($m \times n$ pixels) can be expressed as:

$$I_i = \sum_{x=1}^{m}\sum_{y=1}^{n} S_i(x,y) O(x,y), \qquad (1)$$

where $I_i$ represents the $i$th light-intensity value collected by the detector and $S_i$ represents the $i$th random pattern loaded to the digital micromirror device (DMD). To obtain the reconstructed image of the object, the optimization model based on total variation is expressed as:

$$\begin{aligned} \min \quad & \|c\|_{l_1} \\ s.t. \quad & GO' = c \\ & S'O' = I' \end{aligned} \qquad (2)$$

where $c$ represents the corresponding coefficient vector, $l_1$ represents the $l_1$ norm, $G$ represents the gradient calculation matrix, $S' \in R^{T \times (m \times n)}$ represents $T$ projected random patterns, $O' \in R^{a \times 1}$ represents the object, and $I' \in R^{T \times 1}$ represents the light intensity.

The imaging quality of the reconstructed image is quantitatively compared using the peak signal-to-noise ratio (PSNR) [41] as the evaluation index.

$$\begin{cases} PSNR = 10\log_{10}\dfrac{(2^k - 1)^2}{MSE} \\ MSE = \dfrac{1}{M}\sum_{x,y}(O'(x,y) - O(x,y))^2 \end{cases}, \qquad (3)$$

where MSE is the mean square error, $M$ is the number of pixels in the whole image, and $k$ is the number of bits, which is set to 8.

## 2.1 Temporally variable-resolution illumination patterns

In the conventional approach, obtaining an image with an imaging resolution of $N \times N$ pixels is usually measured using illumination patterns with an actual resolution of $N \times N$ pixels. However, imaging resolution and actual resolution are not equal in all cases. The amount of imaging resolution of the reconstructed image depends on how many cells exist in the illumination pattern. We define the cell as the smallest unit of the illumination patterns in the case of uniform resolution. As shown in Fig. 2, we give an example of a set of four different patterns that have the same actual resolution but different imaging resolutions. Although their actual resolution is $128 \times 128$ pixels, different types of illumination patterns will obtain reconstructed patterns of different imaging resolutions because the sizes of their cells differ and there are only two cases of '0' and '1' inside each cell. For example, each cell of the $16 \times 16$ cell pattern has $8 \times 8$ pixels, and there are only two cases of '0' and '1' in this cell. This pattern has an actual resolution of $128 \times 128$ pixels and an imaging resolution of $32 \times 32$ cells.

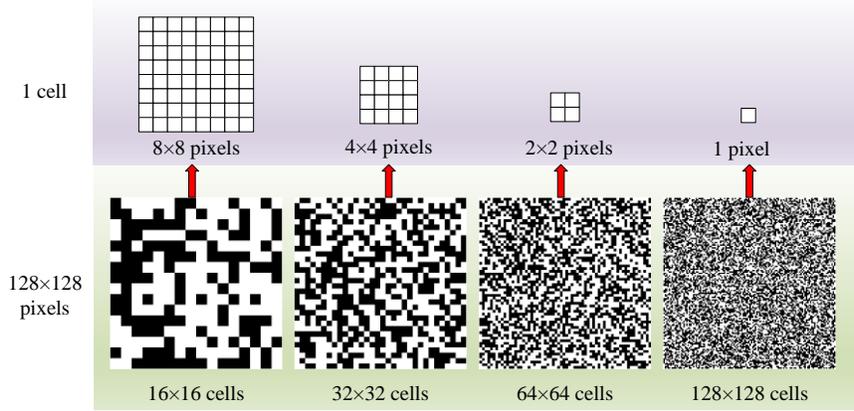

Fig. 2. Four illumination patterns with different imaging resolutions .

We measured and reconstructed the test image 'peppers' using $128 \times 128$ pixel illumination patterns with four different imaging resolutions: $16 \times 16$, $32 \times 32$, $64 \times 64$ and $128 \times 128$ cells. The sampling rate is the ratio of the number of measurements to the actual resolution, where the actual resolution is $128 \times 128$ pixels. The results are shown in Fig. 3. The imaging quality of the reconstructed images using the four different patterns ranges from low to high and then reaches the maximum value of the PSNR. Since the original image for calculating the PSNR is a $128 \times 128$ pixel image, the PSNR of the method measured using $128 \times 128$ cell patterns will always increase. Only the case where the sampling rate is less than 1 is shown in the figure. The differences in the methods that use four different patterns is the moment when they reach the PSNR maximum and the size of the PSNR maximum. It is obvious that the imaging quality of the reconstructed image using $16 \times 16$ cell patterns reaches the PSNR maximum first, but the maximum value of the PSNR is lower than that obtained by the other methods. The highest PSNR maximum was achieved using the patterns with $128 \times 128$ cells, but the last image reached the maximum PSNR. There is a trade-off between imaging efficiency and maximum PSNR. To reach the maximum PNSR at a faster rate, the imaging quality at this time needs improvement. Therefore, we developed temporally variable-resolution illumination patterns to balance the imaging quality and efficiency.

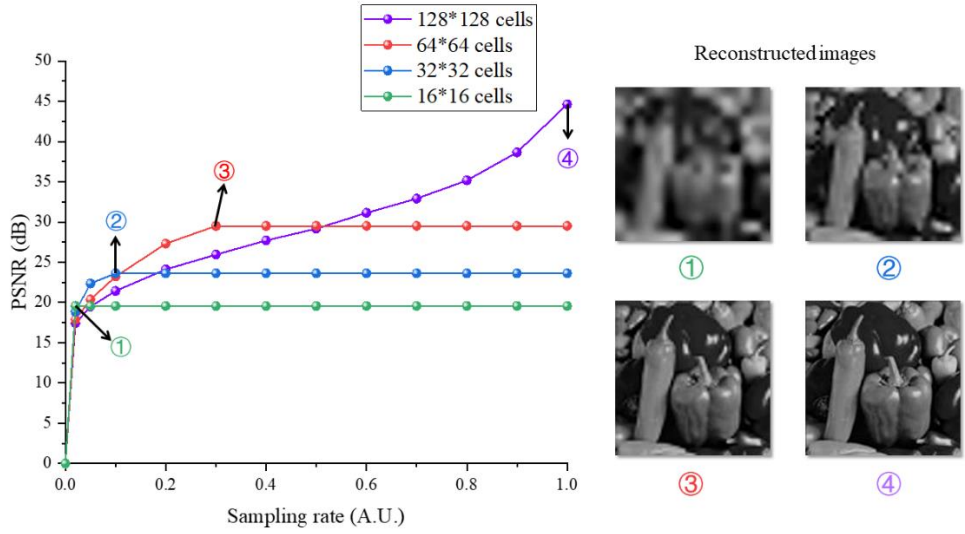

Fig. 3. Reconstructed results of four different methods.

The sequence of patterns that we developed sometimes varies with time. The actual resolution of the patterns is the same, but the imaging resolution is different. As shown in Fig. 4, we project $n_1$ patterns with the lowest imaging resolution of $m_1 \times m_1$ cells. The value of $n_1$ here is $m_1^2$ and is determined by the number of measurements using the patterns with corresponding imaging resolution when the PSNR maximum is reached, since more measurements after the PSNR maximum is reached does not provide any improvement in the performance of the reconstructed image. According to the results of our research, the lowest imaging resolution chosen is 16×16 cells, while $n_1$ should be 256. Second, we project $n_2$ patterns with four times the imaging resolution of the lowest imaging resolution. However, $n_2$ is not $m_2^2$ because $n_1$ measurements have been obtained before projecting $m_2 \times m_2$ cell patterns, so $n_2$ is $m_2^2-n_1$. $m_3$ is twice $m_2$, and $n_3$ is $m_3^2-n_1-n_2$, but $m_k \times m_k$ is not greater than the actual resolution $M \times M$. After measuring a sequence of patterns, the reconstruction calculation is performed as in the CGI. In this way, the number of measurements is equivalent to that in the reconstruction algorithm, and we only need to compare the imaging quality to verify the performance of the illumination patterns that we developed.

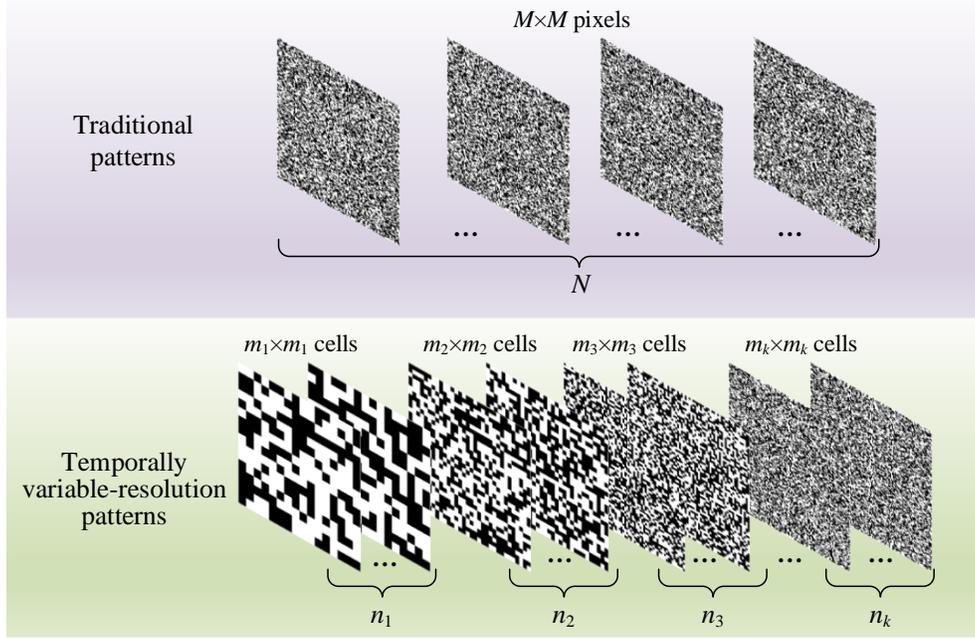

Fig. 4. Sequence of traditional illumination patterns and temporally variable-resolution illumination patterns.

## 2.2 Temporally and spatially variable-resolution illumination patterns

The temporally variable-resolution method solves the problem that it is difficult to achieve high imaging quality with fewer measurements to obtain high-resolution reconstruction images, and the spatially variable-resolution method can improve the imaging quality of the ROI at the expense of information from other nonregions of interest [42]. The goal of both methods is to obtain better imaging quality. We therefore combined these methods and developed TSV illumination patterns. As shown in Fig. 5 (a), (b) and (c), the spatially variable-resolution pattern is a part of the retina-like structure since the pattern that we developed is rectangular and the retina-like structure is toroidal. Both the diameter $2r_0$ of the ROI and the coordinates of the center point ($x_0$, $y_0$) are determined based on the position and size of the object in the reconstructed image. The resolution of the ROI is the highest, and the resolution gradually decreases in the outward direction.

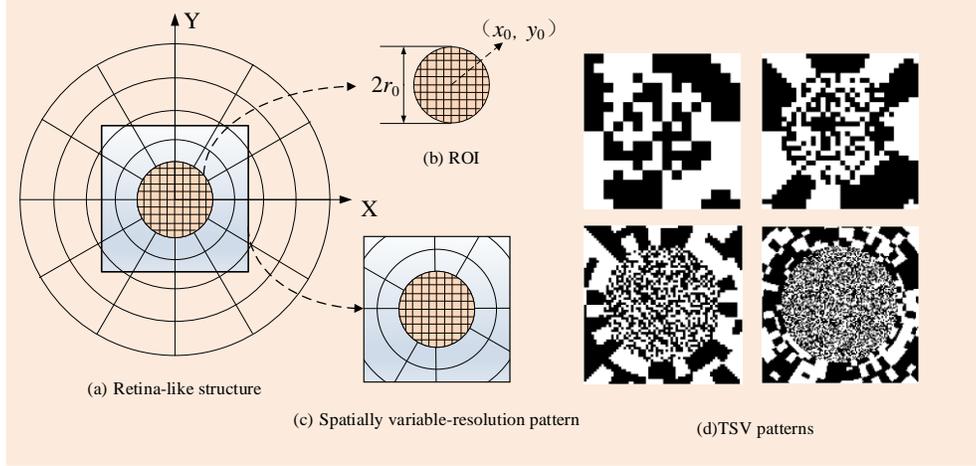

Fig. 5. Generation of TSV illumination patterns. (a) The retina-like structure. (b) The ROI of the spatially variable-resolution pattern. (c) A spatially variable-resolution pattern. (d) Four patterns with different imaging resolutions in their ROI in TSV patterns.

As shown in Fig. 5(d), there are several spatially variable-resolution patterns of 128×128 pixels; in their ROI are 16×16, 32×32, 64×64 and 128×128 cells, respectively. Although their imaging resolutions are different, the coordinates of the center point and the diameter of their ROI are the same. Measurements using the TSV illumination patterns are performed in the same way as those using the temporally variable-resolution patterns. Applying the method that uses TSV illumination patterns can combine the advantages of both variable-resolution methods to achieve better imaging quality with fewer measurements.

## 3. Simulations and experiments

### 3.1 Simulations

We perform the simulation without noise to obtain a reconstructed image of 128×128 pixels. The illumination patterns used for projection are selected as conventional uniform-resolution patterns and temporally variable-resolution patterns. The measurements are simulated by the test image "peppers" at different sampling times, 500, 1000, 2000, 3000, 4000, 5000. The lowest imaging resolution of TVCGI is set as 16×16 cells. The PSNR is selected to evaluate the imaging quality for quantitative comparison. The reconstruction results of "peppers" are shown in Fig. 6. The imaging quality of both methods increases with the number of measurements. However, the imaging quality of TVCGI is superior to UCGI, both subjectively and in terms of PSNR values. With fewer measurements, TVCGI uses patterns with lower imaging resolution for sampling to quickly obtain object outline information, while UCGI uses high-resolution pattern sampling to obtain more detailed information about the object. In most cases, we prefer to view the object and then observe exactly how the object performs. Therefore, TVCGI, which samples by low imaging resolution and then continuously increases the imaging resolution for sampling, is more suitable than UCGI for scenes that require rapid acquisition of object information.

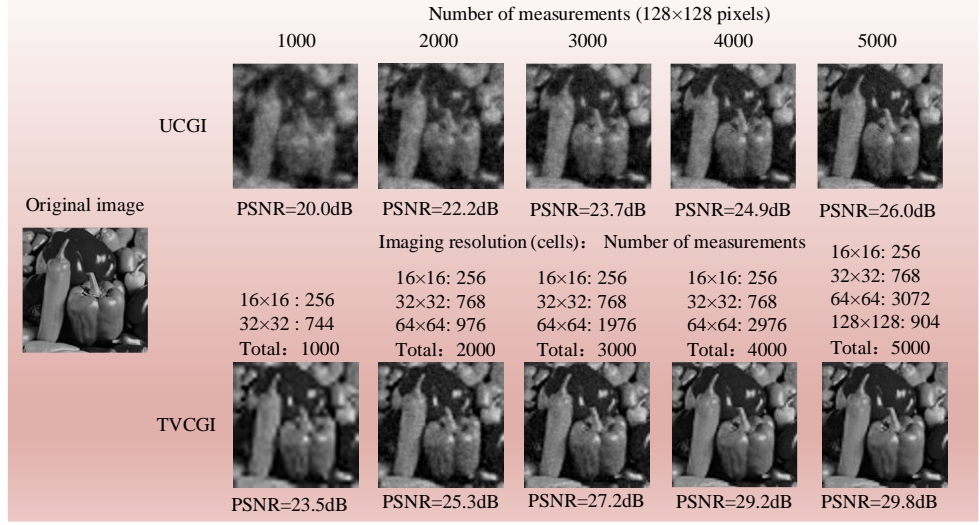

Fig. 6. Comparison results of "peppers" test images simulated by UCGI and TVCGI with different numbers of measurements.

Based on the previous simulations, we choose temporally variable-resolution patterns, spatially variable-resolution patterns, and TSV patterns as the illumination patterns for comparison. For the methods using these three different patterns, we chose the imaging quality of the ROI for comparison. The comparison results are shown in Fig. 7. It is obvious that the imaging quality of the TSVCGI in the ROI is better than that of the TVCGI. However, the imaging quality of the TSVCGI is better than that of the SVCGI when the number of measurements is below 3000 and worse than that of the SVGI when it exceeds 3000. TVCGI has the advantage of quickly obtaining object profile information with fewer measurements and therefore more rapidly improving the imaging quality. However, in the case of numerous measurements, higher imaging quality can be obtained by the method that uses higher imaging resolution patterns. We conclude that better imaging quality can be obtained with the same number of measurements using either the temporally variable-resolution patterns or spatially variable-resolution patterns, and their combination can achieve better performance.

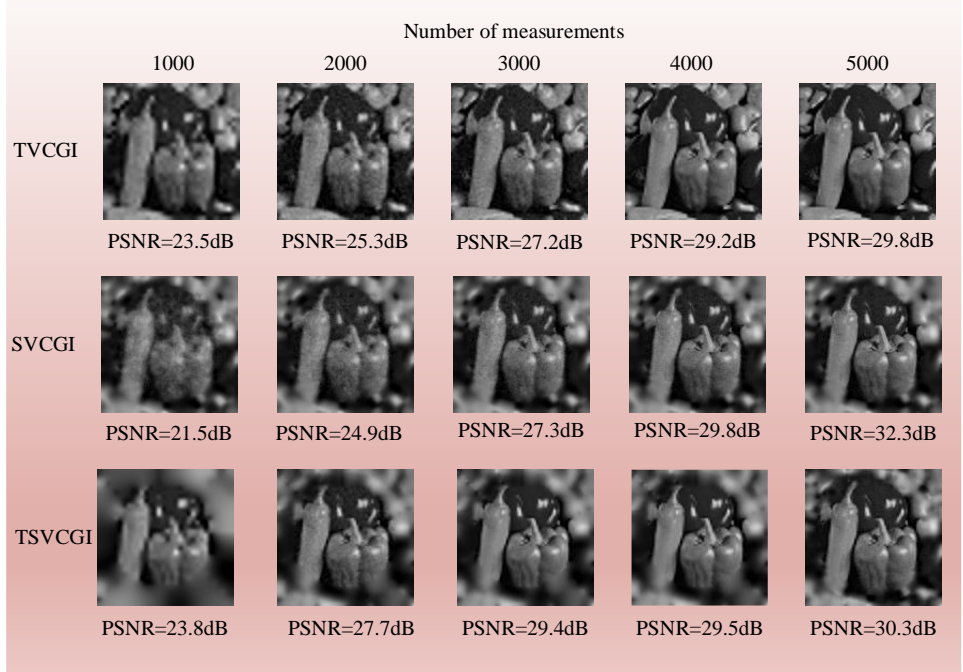

Fig. 7. Comparison results of the "peppers" test image simulated by TVCGI, by computational ghost imaging using spatially variable-resolution patterns (SVCGI) and by TSVCGI with different numbers of measurements.

In a practical optical imaging system, measurements are always affected by noise from ambient light and circuit currents. Noise was not considered in the reconstruction in the above simulation. Simulations on the influence of measurement noise are performed, and the robustness of different methods to noise is analyzed. Here, we assume that white Gaussian noise follows a probability distribution

$$P(c) = \frac{1}{\sqrt{2\pi}\sigma} exp\left(-\frac{c-\mu^2}{2\sigma^2}\right), \quad (4)$$

where $n$ is the noise, $\mu$ is the average value, and $\sigma$ is the standard deviation. We set $\mu$ to 0 and set $\sigma$ to 0, 1, 2, and 3. The number of measurements was set to 5000. The imaging quality of the ROI is employed for comparison. The corresponding reconstructed results are shown in Fig. 8, indicating that the imaging quality of the TSVCGI is better than that of the other two methods when the standard deviation of noise is below 3. It is obvious that UCGI is more robust to noise than TSVCGI and that TSVCGI is more robust to noise than TVCGI. The introduction of the spatially variable-resolution structure improves the robustness of TVCGI to noise but is still more affected by noise than UCGI.

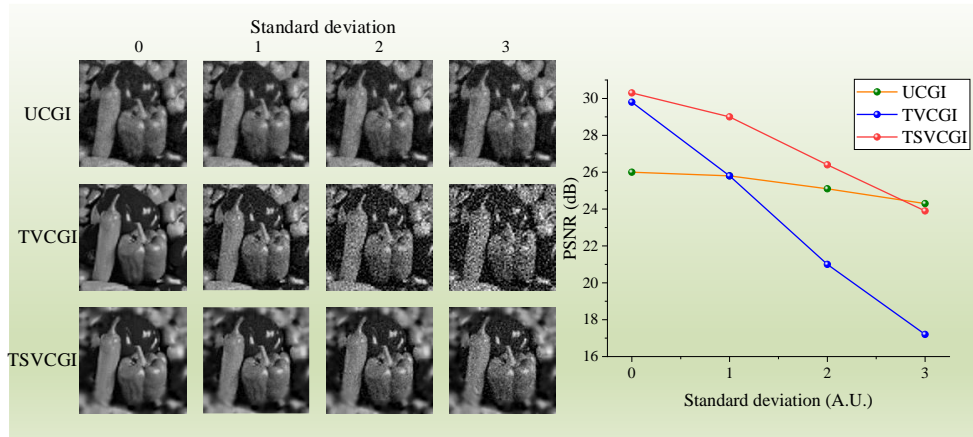

Fig. 8. Comparison results of "peppers" test images simulated by UCGI, TVCGI and TSVCGI with different standard deviations of noise.

### 3.2 Experiments

Based on the simulation results, we further carry out experiments to show the advantages of our methods. A schematic of the experimental setup is shown in Fig. 9. The setup consists of three parts: the illumination part, detection part and object. The illumination part includes a light-emitting diode light source (operates at 400–760 nm, @20 W), DMD (Texas Instruments DLP Discovery 4100 development kit, 1024×768 resolution), and a lens. The maximum binary modulation rate of DMD is 22 kHz. The focal length of the projection lens is 120 mm. The detection part consists of a single-pixel detector (Thorlabs PDA36A, active area of 13 mm$^2$), lens and computer. The focal length of the focusing lens is 50 mm. The object is a "BIT" picture with white letters on a black background. Our experiments were conducted in an environment with no external light source, but noise from ambient light still exists. The noise of this ambient light depends on the working power of the light-emitting diode light source, and the higher the power is, the greater the noise.

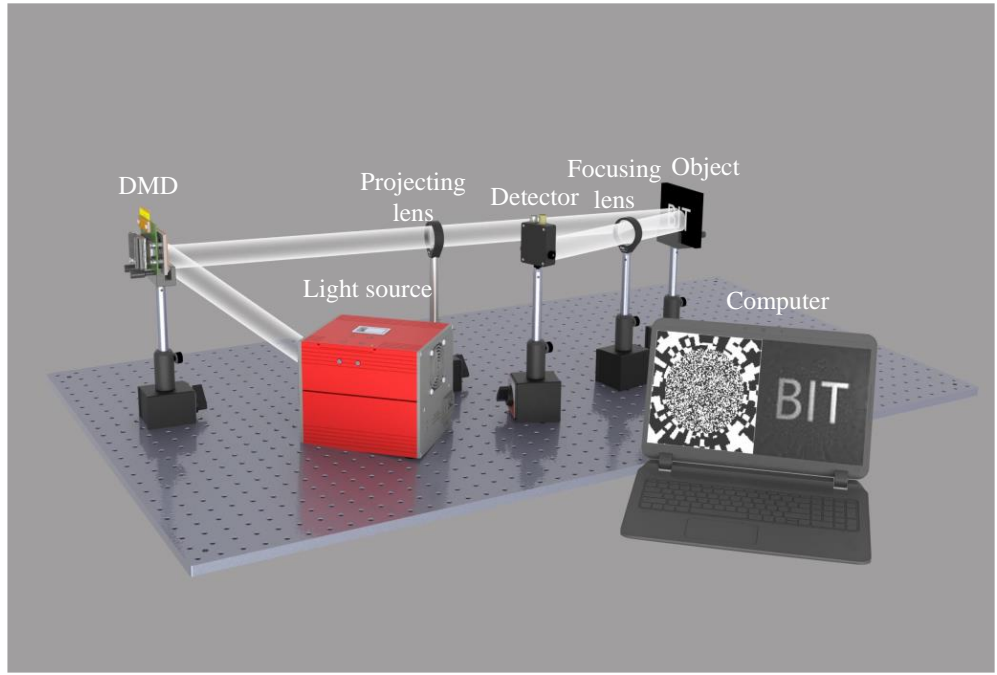

Fig. 9. Schematic of the experimental setup.

We set the number of measurements to 1000, 2000, 3000, 4000 and 5000. The actual resolution of the illumination patterns is 128×128 pixels, and the illumination patterns for projection consist of conventional uniform-resolution patterns, temporally variable-resolution patterns, and TSV patterns. The experimental results are shown in Fig. 10, and we can observe that the imaging quality of the TSVCGI is always better than that of the UCGI and TVCGI. The imaging quality of the TVCGI appears better than that of the UCGI, but the image information is drowned out by noise. For quantitative analysis of the experimental results, we selected the PSNR of the ROI for comparison. The quantitative comparison results are shown in Fig. 11. It is obvious that the imaging quality of the TVSGI in the ROI is better than that of the other methods. The PSNR of both the TVSGI and UCGI in the ROI increases with the number of measurements, but that of the TVCGI increases and then decreases. The reason is that TVCGI has the worst robustness to noise, so the PSNR of TVCGI is higher than UCGI before 4000 measurements, and the PSNR of TVCGI is lower than UCGI after 4000 measurements. It can be concluded that combining the spatially variable-resolution structure with the temporally variable-resolution patterns improves not only the imaging quality of ROI but also the robustness to noise.

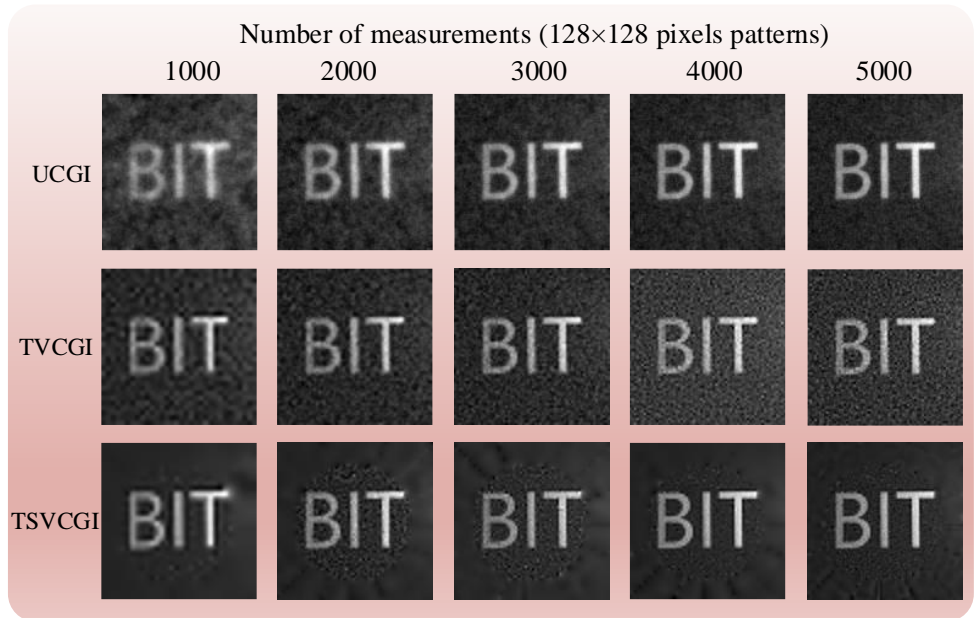

Fig. 10. Experimental results of "BIT" image reconstruction by UCGI, TVCGI, and TSVCGI with different numbers of measurements.

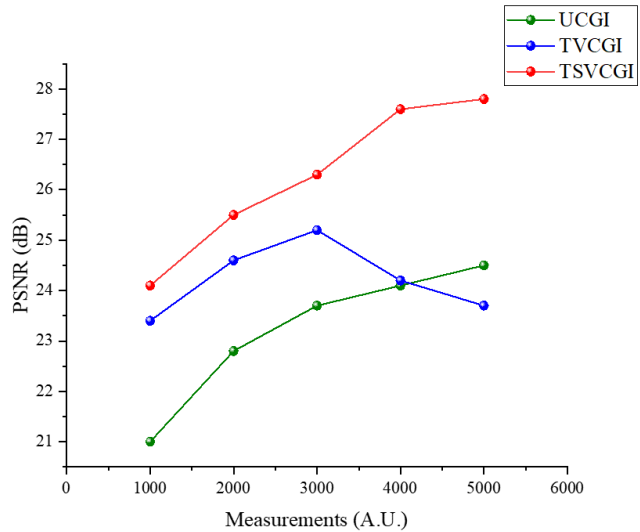

Fig. 11. Quantitative comparison results of "BIT" image reconstruction by UCGI, TVCGI, and TSVCGI with different numbers of measurements.

To verify the performance of the proposed method in performing higher resolution imaging, we generated higher actual resolution patterns for experimental comparison. The actual resolution of the illumination patterns is set to 256×256 pixels, and the illumination patterns for projection are conventional uniform-resolution patterns and TSV patterns. The lowest imaging resolution of the TSVCGI is set to 32×32 cells. The number of measurements is set to 1000, 2000, 3000, 4000, and 5000. The reconstructed results are shown in Fig. 12. It is obvious that the PSNR of the TSVCGI is higher than that of the UCGI in the ROI, despite the influence of ambient noise. The results again verify that our proposed illumination patterns achieve better performance than the method using conventional uniform resolution patterns.

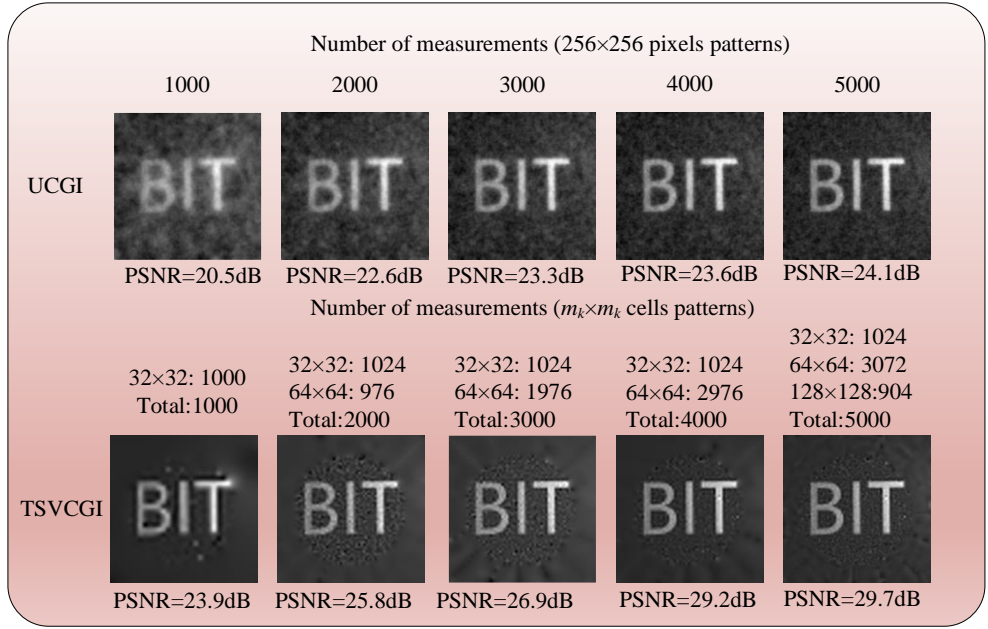

Fig. 12. Experimental results of "BIT" image reconstruction by UCGI and TSVCGI with different numbers of measurements.

We compared the above two groups of experimental results using different actual resolution patterns according to the difference in imaging quality between TSVCGI and UCGI. The comparison results of the difference in imaging quality are shown in Table 1. It is obvious that the improvement in the PSNR of the TSVCGI in the ROI compared with the UCGI in the ROI is better when using patterns with 256×256 pixels than when using patterns with 128×128 pixels. The main reason is that methods with larger actual resolution patterns reach the maximum PSNR value at a lower rate. When the number of measurements is small, using patterns with low imaging resolution can obtain more contour information that contributes more to the improvement in imaging quality. Another reason is that the introduction to the spatially variable-resolution structure also sacrifices the information of nonregions of interest and improves the imaging quality of the ROI. According to the characteristics of our method for obtaining higher image quality with few measurements, it will be well applied for scenes that require fast imaging speed but do not need very high image quality, such as fast high-resolution imaging, fast target tracking, etc.

**Table 1. Quantitative comparison of the PSNR of TSVCGI in ROI minus that of UCGI**

| | Actual resolution of patterns (pixels) | Number of Measurements | | | | |
|---|---|---|---|---|---|---|
| | | 1000 | 2000 | 3000 | 4000 | 5000 |
| ΔPSNR(dB) | 128×128 | 3.1 | 2.7 | 2.6 | 3.1 | 3.3 |
| | 256×256 | 3.4 | 3.2 | 3.6 | 5.6 | 5.6 |

## 4. Discussion and Conclusions

Most CGI is measured using a sequence of illumination patterns with the same imaging resolution, which can make it difficult to balance imaging quality and imaging efficiency when performing imaging to obtain high-resolution images. In this paper, we propose the application of the temporal variable-resolution approach and spatially variable-resolution structure to the design of illumination patterns in CGI. Based on this, we propose temporal variable-resolution patterns and TSV patterns. The proposed illumination patterns are verified by simulations and experiments, and the results prove that TVCGI and TSVCGI can obtain better imaging quality with the same number of measurements than UCGI. In addition, the imaging quality of the TSVCGI in the ROI is better than that of the TVCGI. It was experimentally verified that TSVCGI has better performance in imaging to obtain higher resolution images. We also analyze the influence of noise on the proposed method. The results show that TVCGI has the worst robustness to noise and that TSVCGI is more robust to noise than TVCGI but still inferior to UCGI, which is a shortcoming of our proposed method. This shortcoming can be improved if differential measurement is added to the optical system. The proposed method is expected to solve the current difficulties of CGI to achieve high-quality imaging of high-resolution images.

**Funding.** Beijing Natural Science Foundation (4222017); National Natural Science Foundation of China (61871031, 61875012).

**Acknowledgments.** The authors thank the editor and the anonymous reviewers for their valuable suggestions.

**Disclosures.** The authors declare no conflicts of interest.

**Data availability.** Data underlying the results presented in this paper are not publicly available at this time but may be obtained from the authors upon reasonable request.